\documentclass[useAMS,usenatbib]{mn2e}
\usepackage{txfonts}
\usepackage{graphicx}
\usepackage[mathscr]{eucal}
\newcommand{\be}{\begin{equation}}
\newcommand{\ee}{\end{equation}}
\newcommand{\bdm}{\begin{displaymath}}
\newcommand{\edm}{\end{displaymath}}
\def\dmf{\dot{\mathfrak{M}}}
\def\msE{\mathscr{E}}
\title[Magnetic accretion in SXP\,1062]
{Signs of magnetic accretion in young Be/X-ray pulsar SXP\,1062}
\author[N.R.\,Ikhsanov]{N.R.\,Ikhsanov\\
Central Astronomical Observatory of the Russian Academy of Sciences at Pulkovo,
196140 St.\,Petersburg, Russia}

\begin{document}
\date{Accepted 2012 May 10; Received 2012 April 25}
\pagerange{\pageref{firstpage}--\pageref{lastpage}}
\pubyear{2012}
\maketitle
\label{firstpage}
\begin{abstract}
The spin behaviour of the neutron star in the newly discovered young Be/X-ray long-period pulsar SXP\,1062 is discussed. The star is observed to rotate with the period of 1062\,s, and spin-down at the rate $\sim - 2.6 \times 10^{-12}\,{\rm Hz\,s^{-1}}$. I show that all of the conventional  accretion scenarios encounter major difficulties explaining the rapid spin-down of the pulsar. These difficulties can be, however, avoided within the magnetic accretion scenario in which the neutron star is assumed to accrete from a magnetized wind. The spin-down rate of the pulsar can be explained within this scenario provided the surface magnetic field of the neutron star is $B_* \sim 4 \times 10^{13}$\,G. I show that the age of the pulsar in this case lies in the rage $(2-4) \times 10^4$\,yr, which is consistent with observations. The spin evolution of the pulsar is briefly discussed.
\end{abstract}

\section{Introduction}

A new persistent long-period Be/X-ray pulsar SXP\,1062 has recently been discovered in Small Magellanic Cloud. It has been identified with an accreting magnetized neutron star rotating with the period $P_0 = 1062$\,s. The neutron star is a member of a High Mass X-ray Binary (HMXB) with the orbital period $P_{\rm orb} \simeq 300$\,d. The normal companion is a B0\,IIIe+ star, which underfills its Roche lobe and losses material in a form of the stellar wind. The X-ray luminosity of the pulsar is $L_{\rm X} \simeq 6 \times 10^{35}\,{\rm erg\,s^{-1}}$ \citep{Henault-Brunet-etal-2012, Haberl-etal-2012}.

There are two additional properties which makes SXP\,1062 standing out among presently known Long Period X-ray Pulsars (LPXPs). First, the source has been associated with a supernova remnant of the age $\tau_0 \sim (1-4) \times 10^4$\,yr \citep{Henault-Brunet-etal-2012, Haberl-etal-2012}. The second, the neutron star has been observed to spin-down at the rate $\dot{\nu}_0 \simeq -2.6 \times 10^{-12}\,{\rm Hz\,s^{-1}}$ \citep{Haberl-etal-2012}.

I show in the next section that currently used accretion scenarios encounter major difficulties explaining the rapid spin-down of the pulsar. In order to solve this problem I address the magnetic accretion scenario initially suggested by \citet{Shvartsman-1971} and elaborated analytically by \citet{Bisnovatyi-Kogan-Ruzmaikin-1974, Bisnovatyi-Kogan-Ruzmaikin-1976} and numerically by \citet{Igumenshchev-etal-2003}. The basic ideas of this scenario are briefly outlined in Sect.\,\ref{ma} and its application to SXP\,1062 is discussed in Sect\,\ref{sxp}. I find that the observed spin-down rate of the pulsar can be fitted in this scenario provided the surface magnetic field of the neutron star is $B_* \sim 4 \times 10^{13}$\,G. Possible spin evolution within this model is presented in Sect.\,\ref{evol} and basic conclusions are summarized in Sect.\,\ref{conclusions}.

 \section{Spin-down problem}

The observed spin-down rate of the neutron star in SXP\,1062 indicates that the current spin-down torque applied to the neutron star is limited to $|K_{\rm sd}| \geq 2 \pi I |\dot{\nu}_0|$. Here $I$ is the moment of inertia of the neutron star. If the neutron star were accreting material from a Keplerian disc \citep{Lynden-Bell-Pringle-1974} or from the free-falling spherical flow \citep{Lipunov-1982} the expected spin-down torque would be $K_{\rm sd}^{(0)} \leq \mu^2/r_{\rm cor}^3$. Here $\mu$ is the dipole magnetic moment and
 \be\label{rcor}
r_{\rm cor} = \left(\frac{GM_{\rm ns}}{\omega_{\rm s}^2}\right)^{1/3} \simeq 1.8 \times 10^{10}\ m^{1/3}\ \left(\frac{P_{\rm s}}{P_0}\right)^{2/3}\ {\rm cm}
 \ee
is the corotation radius of the neutron star. $m = M_{\rm ns}/1.4\,{\rm M_{\sun}}$ is the mass of the neutron star and $\omega_{\rm s} = 2\pi/P_{\rm s}$ is its angular velocity.

Solving inequality $|K_{\rm sd}^{(0)}| \geq 2 \pi I |\dot{\nu}_0|$ for $\mu$ one finds
 \be
 \mu \geq 3 \times 10^{32}\ I_{45}^{1/2} m^{1/2} \left(\frac{P_{\rm s}}{P_0}\right)  \left(\frac{|\dot{\nu}|}{|\dot{\nu}_0|}\right)^{1/2}\ {\rm G\,cm^3},
 \ee
where $I_{45} = I/10^{45}\,{\rm g\,cm^2}$, $\dot{\nu} = d\nu/dt$ and $\nu = 1/P_{\rm s}$ is the rotational frequency of the neutron star. This implies the surface field of the neutron star, $B_* = 2 \mu/R_{\rm ns}^3$, to be in excess of $6 \times 10^{14}$\,G (here I assume that the radius of the neutron star is $R_{\rm ns} = 10^6$\,cm). However, the magnetospheric radius of the neutron star under these conditions \citep[see e.g.][]{Arons-Lea-1976, Arons-1993},
    \begin{eqnarray}\label{ra}
 r_{\rm A} & = & \left(\frac{\mu^2}{\dmf \sqrt{2 GM_{\rm ns}}}\right)^{2/7} \simeq 2 \times 10^{10}\,{\rm cm}\ \times\ m^{1/7}\ R_6^{-2/7} \\
  \nonumber
  &  \times & \left(\frac{\mu}{3 \times 10^{32}\,{\rm G\,cm^3}}\right)^{4/7}
  \left(\frac{L_{\rm X}}{6 \times 10^{35}\,{\rm erg\,s^{-1}}}\right)^{-2/7},
    \end{eqnarray}
turns out to be larger that its corotation radius (see Eq.~\ref{rcor}). Here $\dmf = L_{\rm X} R_{\rm ns}/GM_{\rm ns}$ is the mass accretion rate onto the neutron star and $R_6 = R_{\rm ns}/10^6$\,cm. The neutron star in this case is in the centrifugal inhibition (Propeller) state in which no accretion onto its surface occurs \citep[see e.g.][]{Illarionov-Sunyaev-1975, Stella-etal-1986}.

A modification of spherical accretion scenario in which the neutron star is assumed to accrete material from a hot turbulent envelope surrounding its magnetosphere has recently been discussed by \citet{Shakura-etal-2012}. The spin-down torque evaluated within this scenario is limited to $K_{\rm sd}^{\rm (t)} \leq \dmf\,\omega_{\rm s}\,r_{\rm A}^2$. Using the parameters of SXP\,1062 one finds that the condition $|K_{\rm sd}^{\rm (t)}| \geq 2 \pi I |\dot{\nu}_0|$ could be satisfied only if the magnetospheric radius were $r_{\rm A} \geq 3 \times 10^{10}\,{\rm cm}$, which again exceeds the corotation radius.

Thus, an attempt to explain the observed spin-down rate of the neutron star in SXP\,1062 within presently adopted accretion scenarios leads us to a controversy. This controversy may indicate that the conventional accretion picture is oversimplified and does not take into account a factor which under certain conditions strongly affects the accretion process. One of such factors is the magnetic field of the accretion flow. I discuss the role of this factor in the following sections.

  \section{Accretion from a magnetized flow}\label{ma}

A possibility to incorporate the magnetic field of the accretion flow, $B_{\rm f}$, into the spherical accretion model has first been discussed by \citet{Shvartsman-1971}. He has considered a spherical accretion onto a compact star in a wind-fed HMXB. The mass capture rate by the compact star from the stelar wind of its massive companion is $\dmf_{\rm c} = \pi R_{\rm G}^2 \rho_{\infty} V_{\rm rel}$, where $R_{\rm G} = 2 GM_*/V_{\rm rel}^2$ is the Bondi radius, $M_*$ and $V_{\rm rel}$ are the mass and the relative velocity of the compact star, and $\rho_{\infty}$ is the density of the material at its Bondi radius. The captured material is characterized by its ram pressure, $\msE_{\rm ram}(R_{\rm G}) = \rho_{\infty} V_{\rm rel}^2$, thermal pressure, $\msE_{\rm th}(R_{\rm G}) = \rho_{\infty} c_{\rm s}^2$, and magnetic pressure, $\msE_{\rm m}(R_{\rm G}) = B_{\rm f}^2(R_{\rm G})/8 \pi$. The latter can be normalized to the thermal pressure of the material captured at the Bondi radius through the parameter $\beta = \msE_{\rm th}(R_{\rm G})/\msE_{\rm m}(R_{\rm G})$. Here $c_{\rm s}$ is the sound speed in the material captured by the compact star at its Bondi radius. The case $\beta \sim 1$ is referred to as the magnetic accretion.

\citet{Shvartsman-1971} has considered a situation in which the captured material initially follows ballistic trajectories moving towards the compact star with the free-fall velocity, $V_{\rm ff} = \left(GM_*/r\right)^{1/2}$. The magnetic field in the free-falling material is dominated by its radial component, $B_{\rm r}$, \citep[the transverse scales in the free-falling flow contract as $r^{-2}$, while the radial scales expand as $r^{1/2}$,][]{Zeldovich-Shakura-1969}, which under the magnetic flux conservation approximation increases as $B_{\rm r}(r) \propto r^{-2}$ \citep{Bisnovatyi-Kogan-Fridman-1970}. The magnetic pressure in the free-falling material,
\be
 \msE_{\rm m}(r) = \msE_{\rm m}(R_{\rm G}) \left(\frac{R_{\rm G}}{r}\right)^4,
 \ee
increases, therefore, more rapidly than the ram pressure,
\be\label{eram}
 \msE_{\rm ram}(r) = \msE_{\rm ram}(R_{\rm G}) \left(\frac{R_{\rm G}}{r}\right)^{5/2}.
 \ee
The distance at which the condition $\msE_{\rm m}(R_{\rm sh}) = \msE_{\rm ram}(R_{\rm sh})$ is satisfied is \citep{Shvartsman-1971},
\be\label{rsh}
 R_{\rm sh} = \beta^{-2/3} \left(\frac{c_{\rm s}}{V_{\rm rel}}\right)^{4/3} R_{\rm G} =
 \beta^{-2/3}\ \frac{2 GM_*\,c_{\rm s}^{4/3}}{V_{\rm rel}^{10/3}}.
 \ee
This distance in the following consideration is refereed to as the Shvartsman radius.

The accretion process inside Shvartsman radius is fully controlled by the magnetic field of the flow itself and occurs on the timescale of the magnetic field annihilation in the accreting material. Since this time under the conditions of interest significantly exceeds the dynamical (free-fall) time the accretion should switch into a settling stage.

The magnetic accretion scenario has been later elaborated by \citet{Bisnovatyi-Kogan-Ruzmaikin-1974, Bisnovatyi-Kogan-Ruzmaikin-1976} and \citet{Igumenshchev-etal-2003}. These authors confirmed that the magnetic field in the free-falling material is rapidly amplified and plays a key role in the accretion process inside Shvartsman radius. Both of these studies eventually suggest that the accretion flow is decelerated by its own magnetic field at the Shvartsman radius and switches into the settling accretion stage. This transition is accompanied with a change of the flow geometry and formation of a dense slowly rotating slab in which the material is confined by the magnetic field of the accretion flow itself. The material is approaching the compact star as the magnetic field in the slab is annihilating. I take this result as the basis of my following consideration. I assume that the neutron star in SXP\,1062 is accreting material from the magnetic slab which is surrounding its magnetosphere.

  \section{Magnetic accretion in SXP\,1062}\label{sxp}

A necessary condition for the magnetic accretion onto a magnetized neutron star to realize reads $R_{\rm sh} \geq r_{\rm A}$ \citep[for discussion see][]{Ikhsanov-etal-2012, Ikhsanov-Beskrovnaya-2012, Ikhsanov-Finger-2012}. This condition for the parameters of SXP\,1062 can be expressed as $V_{\rm rel} \leq V_{\rm mca}$, where
  \be
 V_{\rm mca} \leq 340\ \beta^{-1/5}\,m^{12/35}\,\mu_{31}^{-6/35}\,c_6^{2/5}\,\dmf_{15.5}^{3/35}\ {\rm km\,s^{-1}}.
 \ee
Here $m$ and $\mu_{31}$ are the mass and dipole magnetic moment in units of $1.4\,{\rm M_{\sun}}$ and $10^{31}\,{\rm G\,cm^3}$, $c_6 = c_{\rm s}/10^6\,{\rm cm\,s^{-1}}$, and $\dmf_{15.5}$ is the mass accretion rate in units of $10^{15.5}\,{\rm g\,s^{-1}}$. The value of $V_{\rm mca}$ is comparable with the relative velocity of the neutron star in SXP\,1062 adopted by \citet{Popov-Turolla-2012}.

If the above condition is satisfied the neutron star accretes material from the magnetic slab. The interaction between the slab and the magnetic field of the neutron star leads to formation of the magnetosphere. The radius of the magnetosphere, $r_{\rm m}$, can be defined by equating the gas pressure in the slab with the pressure due to the dipole magnetic field of the neutron star. This indicates that the gas density of the material at the inner radius of the slab is
 \be\label{rhosl}
  \rho_0 = \frac{\mu^2\,m_{\rm p}}{2 \pi\,k_{\rm B}\,T_0\,r_{\rm m}^6},
  \ee
where $T_0$ is the gas temperature, $m_{\rm p}$ is the proton mass and $k_{\rm B}$ is the Boltzmann constant. The half-thickness of the slab can be approximated following \citet{Bisnovatyi-Kogan-Ruzmaikin-1976} by the height of the homogeneous atmosphere,
 \be\label{hs}
 h_{\rm s}(r_{\rm m}) = \frac{k_{\rm B} T_0 r_{\rm m}^2}{m_{\rm p} GM_{\rm ns}},
 \ee
where $M_{\rm ns}$ is the mass of the neutron star.

The spin-down torque applied to the neutron star from the magnetic slab in the general case can be expressed as
 \be\label{ksdsl}
 K_{\rm sd}^{\rm sl} = S_{\rm eff}\,\nu_{\rm m}\,\rho_0\,V_{\phi}(r_{\rm m}).
 \ee
Here $S_{\rm eff} = 2 \pi r_{\rm m} h_{\rm s}(r_{\rm m})$ is the effective area of interaction between the slab and the magnetosphere and $V_{\phi} = \omega_{\rm s} r_{\rm m}$ is the linear velocity at the magnetospheric boundary (the slab is assumed here to be non-rotating). $\nu_{\rm m} = k_{\rm m} r_{\rm m} V_{\rm A}(r_{\rm m})$ is the magnetic viscosity coefficient and $V_{\rm A} = B_{\rm f}/\sqrt{4 \pi \rho}$ is the Alfv\'en velocity, which in the magnetic slab is equal to the free-fall velocity \citep[for discussion see][]{Bisnovatyi-Kogan-Ruzmaikin-1976}. Finally, $0< k_{\rm m} < 1$ is a dimensionless efficiency parameter.

Combining Eqs.~(\ref{rhosl}) - (\ref{ksdsl}) one finds the maximum value of the spin-down torque which can be applied to a neutron star accreting material from the magnetic slab as
  \be\label{ksdsl1}
K_{\rm sd}^{\rm sl} = \frac{k_{\rm m}\,\mu^2\,\omega_{\rm s}}{r_{\rm m}^{3/2} (2 GM_{\rm ns})^{1/2}} = \frac{k_{\rm m}\,\mu^2}{\left(r_{\rm m}\,r_{\rm cor}\right)^{3/2}}.
 \ee
This equation represents a generalized form of the spin-down torque applied to the neutron star from the accreting material. The conventional expression of the spin-down torque, i.e. $\dmf \omega_{\rm s} r_{\rm A}^2$, can be derived from this equation by putting $r_{\rm m} = r_{\rm A}$. It, however, allows us to consider also a situation in which the accreting material approaches the neutron star to a closer distance than the conventional magnetospheric radius $r_{\rm A}$. This situation may occur if the rate of mass entry into the pulsar magnetosphere is smaller than the mass capture rate by the neutron star from its environment. The accreting material in this case accumulates at the magnetospheric boundary and approaches the neutron star to a closer distance as the gas pressure at the magnetospheric boundary increases \citep[see e.g.][]{Lamb-etal-1973, Lipunov-1992}.

There should not be a confusion that the spin-down torque increases with decrease of the magnetospheric radius. The torque is a twisting force caused by the multiplication of a force by a lever arm. The twisting force applied to the magnetospheric boundary from the accreting material scales with the magnetospheric radius as $S_{\rm eff} \nu_{\rm m} \rho_0 \omega_{\rm s} \propto r_{\rm m}^{-5/2}$, while the lever arm is $\propto r_{\rm m}$. Therefore, the torque applied to the neutron star increases as the material being accumulated at the magnetospheric boundary approaches the neutron star to a closer distance as $\propto r_{\rm m}^{-3/2}$. This is reflected by Eq.~(\ref{ksdsl1}).

The distance to which the magnetic slab approaches the neutron star depends on the mode by which the accreting material enters the magnetic field of the neutron star. If the magnetospheric boundary is interchange unstable (i.e. Rayleigh-Taylor or Kevin-Helmholtz instabilities) the magnetospheric radius is $r_{\rm m} = r_{\rm A}$. The spin-down torque in this case would be the same as predicted within the conventional accretion scenarios (see above). One can, however, envisage a situation in which the interchange instabilities of the boundary are suppressed. The process of plasma entry into the magnetosphere in this case is governed by the diffusion and occurs at the rate \citep{Elsner-Lamb-1984}
    \be\label{dmfin-1}
 \dmf_{\rm dif}(r_{\rm m})  \sim  4 \pi r_{\rm m} \delta_{\rm m} \rho_0 V_{\rm ff}(r_{\rm m})
 = 4 \pi r_{\rm m}^{5/4} D_{\rm eff}^{1/2} \rho_0 (2 GM_{\rm ns})^{1/4}.
   \ee
Here $\delta_{\rm m} = \left(\tau_{\rm d}\ D_{\rm eff}\right)^{1/2}$ is the thickness of the diffusion layer at the magnetospheric boundary (magnetopause), $\tau_{\rm d} \sim t_{\rm ff}(r_{\rm m})$ is the diffusion time, which is normalized to the time on which the material being penetrated into the field leaves the magnetopause by free-falling along the magnetospheric field lines and $D_{\rm eff} \sim D_{\rm B} = \alpha_{\rm B} c k_{\rm B} T_0 r_{\rm m}^3/2 e \mu$ is the effective diffusion coefficient, which is normalized to the Bohm diffusion coefficient, following measurements of the rate of plasma penetration into the Earth's magnetosphere \citep[see][and references therein]{Paschmann-2008}. $e$ is the electron charge and $\alpha_{\rm B}$ is the efficiency parameter, which ranges as $0.1-0.25$ \citep{Gosling-etal-1991}. Putting these parameters to Eq.~\ref{dmfin-1} one finds that the diffusion rate of the accreting material into the pulsar magnetosphere scales as $\dmf_{\rm dif} \propto r_{\rm m}^{-13/4}$.

If $\dmf_{\rm dif} < \dmf$ the material accumulates at the inner radius of the magnetic slab. As it occurs the density and, correspondingly, the gas pressure at the inner radius of the slab increases. The slab, therefore, approaches the neutron star to a closer distance at which the magnetic pressure due to the dipole field of the neutron star is sufficient to balance the increased pressure of the accumulated material. Since $\dmf_{\rm dif} \propto r_{\rm m}^{-13/4}$ the diffusion rate of the material into the magnetosphere also increases. Therefore, the process of plasma accumulation  continues up to a moment at which the condition for stationary accretion process, $\dmf_{\rm dif}(r_{\rm mca}) = \dmf_{\rm c}$, is satisfied. The distance of closest approach of the magnetic slab to the neutron star, $r_{\rm mca}$, can be derived by solving equation $\dmf_{\rm dif}(r_{\rm mca}) = L_{\rm X} R_{\rm ns}/GM_{\rm ns}$. Using the parameters of SXP\,1062 one finds
\be\label{rmca}
 r_{\rm mca} \simeq 7 \times 10^8\ \alpha_{0.1}^{2/13}\,\mu_{31}^{6/13}\,T_6^{-2/13}\,m^{5/13}\,R_6^{-4/13}\,L_{35.8}^{-4/13}\ {\rm cm},
 \ee
where $\alpha_{0.1} = \alpha_{\rm B}/0.1$, $T_6 = T_0/10^6$\,K and $L_{35.8} = L_{\rm X}/10^{35.8}\,{\rm erg\,s^{-1}}$. Combining Eqs.~(\ref{ksdsl1}) and (\ref{rmca}) and solving inequality $|K_{\rm sd}^{\rm (sl)}| \geq 2 \pi I |\dot{\nu}_0|$ leads us to a conclusion that the observed spin-down rate of the neutron star in SXP\,1062 can be fitted into the magnetic accretion model provided the surface field of the neutron star is
    \begin{eqnarray}\label{bfinal}
 B_*  & \geq & 4 \times 10^{13}\,{\rm G}\ \times\ k_{\rm m}^{-13/17} \alpha_{0.1}^{3/17} m^{14/17} R_6^{-57/17}\ \times \\
     \nonumber
 &  &  \times\ I_{45}^{13/17}\ T_6^{-3/17}\ L_{35.8}^{-6/17}\ \left(\frac{P_{\rm s}}{P_0}\right)^{13/17}  \left(\frac{|\dot{\nu}|}{|\dot{\nu}_0}\right)^{13/17}.
   \end{eqnarray}

  \section{Possible origin of SXP\,1062}\label{evol}

It is widely adopted that a newly formed neutron star rotates rapidly with a period of a fraction of a second. Its rotational rate then decreases, initially by the spin-powered pulsar mechanism (Ejector state), and later by means of the interaction between the magnetosphere of the neutron star and the accretion flowing within its Bondi radius (Propeller state). As the spin period of the neutron star reaches a critical value the accretion of material onto its surface starts (Accretor state) and the star switches on as an X-ray pulsar \citep[for a review see,  e.g.,][]{Bhattacharya-van-den-Heuvel-1991, Iben-Tutukov-Yungelson-1995}.

According to modern views \citep[see e.g.][and references therein]{Spitkovsky-2006, Beskin-2010} the spin-down power of a neutron star in the Ejector state can be evaluated as  \citep{Spitkovsky-2006}
 \be\label{wej0}
 W_{\rm ej} = \frac{\mu^2 \omega_{\rm s}^4}{c^3} \left(1+\sin^2{\chi}\right),
 \ee
where $\chi$ is the angle between the rotational and magnetic axes of the neutron star (note, that $\mu = (1/2) B_* R_{\rm ns}^3$). The spin-down power in this case is converted into the electromagnetic waves and relativistic particles which are collectively referred to as the relativistic wind.

The neutron star remains in the Ejector state as long as the pressure of the relativistic wind at  the Bondi radius, $p_{\rm w}(R_{\rm G}) = W_{\rm ej}/4 \pi R_{\rm G}^2 c$, dominates the ram pressure of the surrounding material, $\msE_{\rm ram} = \rho_{\infty} V_{\rm rel}^2$. The spin period at which the Ejector spin-down ceases, $P_{\rm ej}$, can be derived by solving equation $p_{\rm w}(R_{\rm G}) = \msE_{\rm ram}$. Assuming the pulsar to be a nearly orthogonal rotator ($\chi \sim 90^{\degr}$), one finds
 \be
 P_{\rm ej} \simeq 0.8\ \mu_{31.3}^{1/2}\,\dmf_{15.5}^{-1/4}\,\left(\frac{V_{\rm rel}}{300\,{\rm km\,s^{-1}}}\right)^{-1/4}\ {\rm s}.
 \ee
The time, which a neutron star spent in the Ejector state, $\tau_{\rm ej} = P_{\rm ej}/2 \dot{P}(P_{\rm ej})$, is
 \be\label{tauej}
 \tau_{\rm ej} \simeq 3 \times 10^4\ I_{45}\,\mu_{31.3}^{-1}\,\dmf_{15.5}^{-1/2}\,\left(\frac{V_{\rm rel}}{300\,{\rm km\,s^{-1}}}\right)^{-1/2}\ {\rm yr}.
 \ee

The transition of the pulsar from the Ejector to Propeller state is associated with the formation of the accretion flow inside the Bondi radius \citep{Davies-Pringle-1981}. This transition within the conventional non-magnetized flow approximation ($\beta \gg 1$) leads to formation of a hot turbulent quasi-stationary atmosphere which is surrounding the magnetosphere of the neutron star. The spin-down timescale of the neutron star in the Propeller state within this approach can be evaluated from Eq.~(21) of \citet{Ikhsanov-2007} and under the conditions of interest (i.e. $\mu = 2 \times 10^{31}\,{\rm G\,cm^3}$, $\dmf = 3 \times 10^{15}\,{\rm g\,s^{-1}}$ and $V_{\rm rel} = 300\,{\rm km\,s^{-1}}$) turns out to be in excess of $5 \times 10^5$\,yr.

If, however, the accretion onto the neutron star is realized according to the magnetic accretion scenario (i.e. $\beta \sim 1$), the pulsar transition into the Propeller state can be accompanied with the formation of the magnetic slab. The material in the slab is approaching the neutron star up to the distance $r_{\rm mca}$ (see Eq.~\ref{rmca}) at which the rate of plasma diffusion into the stellar field reaches the mass capture rate of the neutron star from its environment. As long as $r_{\rm mca} > r_{\rm cor}$ the centrifugal barrier at the magnetospheric boundary prevents the accretion flow from reaching the stellar surface and the material being penetrated into the magnetopause will be ejected \citep[for discussion see e.g.][]{Lovelace-etal-1995, Perna-etal-2006}. The spin-down torque applied to the neutron star in this case can be evaluated as $\sim \mu^2/r_{\rm mca}^3$. This indicates that the time which the neutron star is expected to spend in the Propeller state within the magnetic accretion scenario is
 \be\label{tauprop}
 \tau_{\rm prop} = \frac{\pi I r_{\rm mca}^3}{\mu^2 P_{\rm mca}},
 \ee
where $P_{\rm mca}$ is the spin period at which the neutron star switches from the Propeller to Accretor state. This period can be evaluated by solving equation $r_{\rm mca} = r_{\rm cor}$ as
 \be\label{pmca}
 P_{\rm prop} \sim 12\ \alpha_{0.1}^{3/13}\ \mu_{31.3}^{9/13}\ T_6^{-3/13}\ m^{17/26}\ R_6^{-6/13}\   L_{35.8}^{-6/13}\,{\rm s}.
 \ee
Combining Eqs.~(\ref{rmca}), (\ref{tauprop}) and (\ref{pmca}) one finds
      \be
 \tau_{\rm prop}  \sim\  2300\  \alpha_{0.1}^{9/13} \mu_{31.3}^{1/13} T_6^{-9/13}\ I_{45}\  m^{77/26}  R_6^{-18/13} L_{35.8}^{-18/13}\ {\rm yr}.
     \ee

The spin-down timescale of the neutron star in the Accretor state within the magnetic accretion scenario is limited to
 \be
 \tau_{\rm acc} \leq \frac{1}{2 |\dot{\nu}_0| P_{\rm prop}} \simeq 535\,\left(\frac{|\dot{\nu}|}{|\dot{\nu}_0}\right)^{-1}\ \left(\frac{P_{\rm prop}}{12\,{\rm s}}\right)^{-1}\ {\rm yr}.
 \ee

Thus, our analysis shows that young age ($\sim 3.3 \times 10^4$\,yr) of the long-period pulsar SXP\,1062 can be explained within the magnetic accretion scenario provided its initial surface field was $\sim 4 \times 10^{13}$\,G. This is consistent with the age of the pulsar evaluated by \citet{Henault-Brunet-etal-2012}. According to \citet{Urpin-etal-1998}, the field of this strength does not significantly decay on the time scale of $\la 10^5$\,yr and, therefore, it remains close to its initial value in the current epoch.

A younger age of the pulsar, $\tau \sim (1.6 - 2.5) \times 10^4$\,yr, reported by \citet{Haberl-etal-2012}, can be explained within this scenario if either the initial surface field of the neutron star was $\sim 8 \times 10^{13}$\,G (see Eq.~\ref{tauej}) or the spin-down power of the neutron star in the state of Ejector was in excess of $W_{\rm ej}$ (see Eq.~\ref{wej0}). The second possibility has been previously discussed by \citet{Beskin-etal-1993}, who suggested that the spin-down power of a neutron star in the Ejector state can under certain conditions be a factor of 2 larger than $W_{\rm ej}$ \citep[see Eq.~113 in][]{Beskin-2010}. If this mechanism is realized in SXP\,1062 the pulsar can be as young as $\tau \sim 1.5 \times 10^4$\,yr having been formed with the initial surface field of $B_0 \sim 4 \times 10^{13}$\,G.

Finally, the spin-down timescale of the pulsar observed in the current epoch does not exceed $\left(2P_0 \dot{\nu}_0\right)^{-1} \sim 6$\,yr. This indicates that the observed rapid spin-down is just an occasional fluctuation. It seems quite natural to assume that the pulsar rotates at the equilibrium period, $P_{\rm eq}$, which is defined by equating the time-averaged spin-up and spin-down torques applied to the neutron star from the accretion flow.

The spin-up torque applied to the neutron star in a wind-fed HMXB can be expressed as $K_{\rm su} = \xi\,\Omega_{\rm orb} R_{\rm G}^2 \dmf$ \citep{Illarionov-Sunyaev-1975}, where $\Omega_{\rm orb} = 2 \pi/P_{\rm orb}$ is the orbital angular velocity and $\xi$ is the parameter accounting for dissipation of angular momentum in the accretion flow \citep[see][and references therein]{Ruffert-1999}. The maximum possible value of the spin-up torque applied to the neutron star in SXP\,1062,
 \be
K_{\rm su} \sim 10^{32}\,\xi\,m^2\,\dmf_{15.5}\,\left(\frac{V_{\rm rel}}{300\,{\rm km\,s^{-1}}}\right)^{-4}\ \left(\frac{P_{\rm orb}}{300\,{\rm d}}\right)^{-1}\ {\rm dyne\,cm},
 \ee
is, however, significantly smaller than the current spin-down torque inferred from observations,
 \be
|K_{\rm sd}^{\rm obs}| \ga 2 \pi I |\dot{\nu}_0| \simeq 1.6 \times 10^{34}\ I_{45} \left(\frac{|\dot{\nu}|}{|\dot{\nu}_0|}\right)\ {\rm dyne\,cm}.
 \ee
This indicates that the time-averaged value of the spin-down torque applied to the neutron star in this system is smaller than its current value by a factor of more than 100.

This problem can be avoided if one takes into account rotation of the material in the magnetic slab. The spin-down torque applied to the neutron star from the magnetic slab decreases rapidly as the angular velocity of the material in the inner radius of the slab approaches the angular velocity of the neutron star itself ($V_{\phi} \rightarrow 0$ as $\omega_{\rm sl} \rightarrow \omega_{\rm s}$, see Eq.~\ref{ksdsl}). The angular velocity of the accreting material scales with the radius as $\omega_{\rm sl}(r) \sim \xi\,\Omega_{\rm orb}\,\left(R_{\rm G}/r\right)^2$ \citep{Bisnovatyi-Kogan-1991}. The stable rotation of the neutron star (i.e. $I \dot{\nu} \sim 0$) can, therefore, be expected if $\omega_{\rm sl}(r_{\rm mca}) \sim \omega_{\rm s}$. This condition is satisfied if the time-averaged value of the parameter $\xi$ in the  accreting material is
 \begin{eqnarray}
 <\xi> & \sim & 0.06\ m^{-2}\ \left(\frac{P_{\rm s}}{P_0}\right)^{-1} \left(\frac{P_{\rm orb}}{300\,{\rm d}}\right) \left(\frac{V_{\rm rel}}{300\,{\rm km\,s^{-1}}}\right)^4\ \times\\
  \nonumber
  & & \times\  \left(\frac{r_{\rm mca}}{7 \times 10^8\,{\rm cm}}\right)^2.
 \end{eqnarray}
This is a factor of 3 smaller than the time-averaged value of $\xi$ derived in the numerical modelings of spherical accretion within the non-magnetized flow approximation \citep{Ruffert-1999}.
A higher efficiency of dissipation of angular momentum in the accretion flow can be expected in the accreting material is magnetized \cite[see e.g.][]{Mestel-1959, Sparke-1982}. This finding provides us with an additional hint that the magnetic field of the accreting material should be taken into account in modeling of the accretion process in SXP\,1062.

   \section{Conclusions}\label{conclusions}

SXP\,1062 represents an exceptional case of a hard braking slowly rotating neutron star in a wind-fed HMXB. All currently used accretion scenarios encounter major difficulties explaining the spin-down rate of this accretion-powered pulsar. These difficulties can be avoided, however, if one assumes that the neutron star captures material from a magnetized wind ($\beta \sim 1$). This leads us to the magnetic accretion scenario \citep{Shvartsman-1971} in which the neutron star accretes material from a slowly rotating slab confined by the magnetic field of the accretion flow itself \citep{Bisnovatyi-Kogan-Ruzmaikin-1974, Bisnovatyi-Kogan-Ruzmaikin-1976}. The spin-down rate of the neutron star can be fitted in this scenario provided the surface field of the neutron star is $B_* \sim 4 \times 10^{13}$\,G.

The young age ($\tau \sim (2-4) \times 10^4$\,yr) of the pulsar can be explained in the frame of this model provided the initial surface field of the neutron star was $B_0 \ga 4 \times 10^{13}$\,G. The spin-down track of the pulsar contains three basic states: Ejector, Propeller and Accretor. The spin-down time of the neutron star in the Ejector (spin-powered pulsar) state is $\tau_{\rm ej} \sim (1-3)\times 10^4$\,yr. The duration of the Propeller and Accretor states within the magnetic accretion scenario are $\tau_{\rm prop} \sim (2-3) \times 10^3$\,yr and $\tau_{\rm acc}\sim 500-600$\,ys. If the pulsar rotates at the equilibrium period the time-averaged spin-down torque applied to the neutron star is significantly smaller than the current value of spin-down torque inferred from observations. This can be explained in the picture of magnetic accretion provided the material at the inner radius of the magnetic slab rotates with the angular velocity $\omega_{\rm sl}(r_{\rm mca}) \sim \omega_{\rm s}$. This implies that the time-averaged value of the parameter accounting for the angular momentum dissipation in the accretion flow is $<\xi>\,\sim\,0.06$.

\section*{Acknowledgments}

I would like to thank G.S.\,Bisnovatyi-Kogan, L.A.\,Pustil'nik, N.G.\,Beskrovnaya and V.S.\,Beskin  for useful discussions and suggesting improvements.The research has been supported by the Program of Presidium of Russian Academy of Sciences N\,21, and NSH-1625.2012.2.

\label{lastpage}

\end{document}